\documentclass[10pt]{iopart}
\usepackage{latexsym}
\usepackage{epsfig}
\usepackage{epstopdf}

\usepackage{iopams}  
\begin{document}

\title{On the relationship between Red Rectangle and diffuse interstellar bands}

\author{Fr\'ed\'eric Zagury }

\address{Institut Louis de Broglie, 23 rue Marsoulan, 75012 Paris, France}
\ead{fzagury@wanadoo.fr}
\begin{abstract}
A careful examination of Red Rectangle bands which have been considered as diffuse interstellar bands (DIBs) in emission shows that a few are likely to be artifacts in the spectrum.
Some others result from atmospheric extinction.
Consequences for the Red Rectangle band/DIB associations are examined.

I will also comment a striking resemblance between the DIB spectrum and the spectrum of NO$_2$ in the 6150-6250~\AA\ region.
This suggests that some DIBs could be provoked by atmospheric molecules.

\end{abstract}
\vspace{2pc}
\maketitle

\section{Introduction}
The spectrum of the Red Rectangle nebula consists in a hierarchy of interwoven structures.
The broad ERE (Extended Red Emission) bump, responsible for the red color of the nebula, extends over more than 2000~\AA\ between 5500~\AA\ and 7500~\AA.
On the bump, with a resolution of a few \AA, Schmidt et al.  \cite{schmidt80} distinguishes six separate sets of emission-like structures (see also \cite{warren81}).
An increase of the resolution by a factor of ten (a few 0.1~\AA, \cite{winckel02}) reveals fine structure in the three largest Schmidt et al. bands and new features on the bump.

At the beginning of the 1990s' correlations were found \cite{fossey90, sarre91} between prominent Red Rectangle  bands ($\lambda\lambda$5799, 5855 in the $5800\,\rm\AA$ complex, $\lambda6380$, $\lambda6617$), and the position in the spectrum of  important DIBs  ($\lambda\lambda$5797, 5850, 6379, 6614).
Scarrott \emph{et al.}, Sarre \emph{et al.} \cite{scarrott92, sarre95} added a few such Red Rectangle band/DIB associations.
These relationships first suggested that the same carrier was seen in absorption in the direction of reddened stars and in emission in the Red Rectangle.

However, peak wavelengths of the Red Rectangle bands do not coincide with DIB wavelengths.
For the red-degraded (with a steep blue edge and a long red tail) Red Rectangle bands, peak wavelengths are not at a fixed position.
They progressively shift to the blue when distance from HD44179 (the star at the center of the Red Rectangle which illuminates the nebula) increases, due to a narrowing of their red wings \cite{winckel02, scarrott92, sarre95,schmidt91}.
Blue edges are stable and remain along the red side of the corresponding DIBs.
It is thus the blue band head of a red-degraded Red Rectangle band that matches the corresponding DIB wavelength \cite{herbig95}.
Van Winckel  \emph{et al.}, Glinski \& Anderson \cite{winckel02,glinski02} definitely proved that Red Rectangle bands' central wavelengths never reach DIB wavelengths.
 
For the symmetric $\lambda6378$ Red Rectangle band, DIBs have been found on each of its sides ($\lambda\lambda6376$ and 6379)  \cite{scarrott92, herbig95}.

The non-convergence of Red Rectangle bands to DIB wavelengths led \cite{glinski02} to question the DIB/Red Rectangle band relationships.
The very large number of DIBs known today (over 380 in the recent  Hobbs et al. survey \cite{hobbs08}, against less than 80 at the time the first DIB/Red Rectangle band relationships were established), can indeed favor accidental coincidences.

The relationship, however, not only relies  on wavelength positions, but also on physical parameters such as  bands' strength and shape (\cite{scarrott92, herbig95} and references therein).
The 5800~\AA\ region for instance contains among the most important DIBs and the most spectacular Red Rectangle bands, which of course accredits the idea of a relation between the two.

In  two previous studies \cite{rr,rrb} on the Red Rectangle I have outline the importance atmospheric extinction (more specifically: atmospheric absorption and starlight diffracted or scattered in the atmosphere) might have in the observed spectrum of the nebula.
This past work may contribute to the discussion on the  DIB/Red Rectangle band associations in two ways.

It should first be remarked that the positions of  DIB wavelengths on the sides of Red Rectangle bands, if meaningful, characterize an absorption spectrum.
In an absorption spectrum,  it is the minima  in the spectrum which correspond to absorption lines' positions and have physical meaning, as it is observed in the Red Rectangle nebula.

I have also found \cite{rrb} that the sharp $\lambda6378$ and the diffuse  $\lambda6225$  Schmidt et al. bands are present in the background of the Red Rectangle nebula observations.
As mentioned above, $\lambda6378$ is in-between two DIBs.
According to table~5 in  \cite{winckel02} correspondences can be found between sub-structures in $\lambda6225$, and DIBs in the $\lambda6200$ complex \cite{chlewicki87}.
The presence of the two bands in background spectra  either questions their relation to DIBs, or has implications which need to be investigated: can there be an atmospheric origin to some DIBs; what alternative is there?
 
In the present study, I will first re-examine a few Red Rectangle bands which have been associated to DIBs (Sect.~\ref{ana}; Sects.~\ref{data} and \ref{rrselect} present the data).
Results of this analysis, the problems they raise  and hypotheses they suggest, are discussed in Sect.~\ref{dis}.

Different carriers are proposed for the DIBs but no laboratory spectrum which could support a concrete identification  has ever been presented. 
In Sect.~\ref{dib6200}, I will comment on the striking resemblance I have found between the DIB spectrum in the 6200~\AA\ region (related to Red Rectangle $\lambda6225$ diffuse band) and the spectrum of NO$_2$.
\section{Data} \label{data}
Most of the observations used in this work come from \cite{rr,rrb}.

Red Rectangle data consist in three observations retrieved from the ESO Archive Facility, and one observation of the nebula observed at the Fred L. Whipple Observatory   with the FAST spectrograph \cite{fabricant}.
Main characteristics of these data are summarized in Table~\ref{tbl:0}.

As in my previous articles, the spectrum at pixel `$x$' in the 2-D array of one of these long-slit observations is noted `$sx$'.
Figures present raw data, flat-fielded but not corrected for atmospheric extinction and with no background  subtracted.
 
A long-slit ESO data-set consists in 600 spectra. 
For the two observations $11''$ from HD44179 the nebula extends roughly from $s330$ to $s429$  ($\sim 27''$ on the sky, \cite{rrb}).
Maximum brightness in the nebula is reached for spectra close to $s360$.

The FAST observation has 120 pixels. 
The nebula extends from $\sim s38$ to $\sim s63$ ($\sim 31''$).

In Sect.~\ref{dib6200} I will also use  spectra with comparable resolutions of BD+40$^{\circ}$4220  (R=45000, $\Delta\lambda\sim 0.03\,\rm\AA$, \cite{gala00}),  provided by  G.~Galazutdinov, and  of NO$_2$ (laboratory spectrum at room temperature, pressure $\sim10 ^{-3}$~torr, res.=0.01~\AA), due to A. Jenouvrier (Universit\'e de Reims, France).

The latter spectrum is a private communication from A.~Jenouvrier.
It was initially designed for studies of the atmosphere and has served for the HITRAN (HIgh-resolution TRANsmission molecular absorption;  http://cfa-www.harvard.edu/hitran/) database. 
\begin{table}[]
\caption[]{Red Rectangle observations}		
\footnotesize 
    \begin{tabular}{|c|c|ccccc|c|cc|c|}
\hline
&Observatory &Spectro.&Slit$^{(1)}$ & $\Delta \Theta$$^{(2)}$ &$\Delta\lambda$$^{(3)}$ & $\lambda_{inf}-\lambda_{sup}$$^{(4)}$ & d$^{(5)}$ & Data.$^{(6)}$&Obs. date & Ref.$^{(9)}$\\
\hline
1&F. L. Whipple     & FAST & $3''\times3'$ & $1.2$  &1.4    &3660-7530 & 14& -&2001.12.22&\cite{rrb}\\
2$^{(7)}$&La Silla Paranal & EMMI &$1''\times3'$ &$0.268$ & 0.32 &5540-6195 & 6  & 01:07:07.550 &1998.01.26&\cite{winckel02,rrb}\\
3$^{(8)}$&La Silla Paranal & EMMI &$1''\times3'$ &$0.268$ & 0.64 &5520-6880 & 11& 01:58:37.630 &1998.01.26&\cite{winckel02,rrb}\\
4$^{(8)}$&La Silla Paranal & EMMI &$1''\times3'$ &$0.268$ & 0.64 &5520-6880 & 11& 03:01:24.630 &1998.01.26 &\cite{winckel02,rrb}\\
\hline
\end{tabular} 
   \normalsize
\begin{list}{}{}
\item[$(1)$] Slit dimensions on the sky.
\item[$(2)$]Spatial resolution ($''$).
\item[$(3)$]Spectral resolution (\AA).
\item[$(4)$]Wavelength coverage.
\item[$(5)$]Distance  ($''$) from HD44179 (for all observations the slit is roughly perpendicular to the north-south axis of the nebula). 
\item[$(6)$]Spectrum's designation in ESO archive. Should be preceded by 'ONTT.1998-01-26T'.
\item[$(7)$]Cut C in  \cite{winckel02}, fig.~1.
\item[$(8)$]Cut D, in \cite{winckel02}, fig.~1.
\item[$(9)$]Reference papers.
\end{list}
\label{tbl:0}
\end{table}
\begin{table*}
\caption[]{DIBs' related Red Rectangle bands$^{(1)}$}		
       \[
    \begin{tabular}{|c|c|c|c|c|}
\hline
$\lambda_c^{(2)} $ &$\lambda_{inf}^{(3)}$&$\lambda_{sup}^{(4)}$&DIBs$^{(5)}$& Origin$^{(6)}$ \\
\hline
5768  &   5762& 5775& 5760, 5763, 5766, 5769, 5776 & bga\\
5799 &   5790 &  5810 &5797, 5807, 5809, 5812 & n\\
5853   & 5850 & 5865 & 5850, 5854, 5856 & n\\
6197   & ? &6200 & 6196, 6199 & ?  \\
6204   & 6201 &6209 & 6199, 6203, 6205, 6212  & bga \\
6221   & 6215 & 6230 & 6216, 6221, 6224, 6226, 6234 & bga \\
6235   & 6232 & 6238 & 6234, 6237 & bge \\
6446   &6441  &  6449 & 6239, 6445 & ? \\
6615   & 6610 &  6621 & 6614 & n \\
6711   &6709  & 6714 & 6709 & ?\\

\hline
\end{tabular} 
 \]
\begin{list}{}{}
\item[$(1)$]From table~5 in \cite{winckel02}, to which I have added $\lambda6221$.  
\item[$(2)$] Estimated peak wavelength of the band. All wavelengths are given in \AA\ and round up to the nearest integer.
\item[$(3)$]Estimated blue limit of the band.
\item[$(4)$]Estimated red end of the band.
\item[$(5)$]DIBs in the $[\lambda_{inf},\,\lambda_{sup}]$ region (from \cite{gala00}).
\item[$(6)$]'bg' if the line is in the background, 'e' if it is a telluric emission line ('a' otherwise), 'n' when no evidence for an atmospheric origin, '?' when the feature is not certain. 
\end{list}
\label{tbl:1}
\end{table*}
\section{Red Rectangle bands selection} \label{rrselect}
Van Winckel \emph{et al.} [3] (table~5) establish a link between a few Red Rectangle bands and DIBs. 
My purpose in this study is to re-analyse  the origin of these bands. 
First column  of Table~\ref{tbl:1}  gives the peak wavelength of each band, second and third the estimated limits on each of their sides.
$\lambda 6221$ (table~3 in \cite{winckel02}) was added, because of its importance in the $\lambda6225$ complex (Sect.~\ref{l6225}).

The list does not intend to be exhaustive.
A comparison of table~3 in \cite{winckel02} to the DIB catalog of  \cite{gala00} proves that most, if not all, Red Rectangle bands have possible DIB connections.

In the 5800~\AA\ region (from 5790~\AA\ to 5950~\AA) a few other DIB/Red Rectangle band associations can be found in \cite{scarrott92, sarre95}.
General remarks on this particularly complex region of the spectrum are given in Sect.~\ref{l5800}.

At longer wavelengths, six out of the eleven Red Rectangle bands listed in \cite{scarrott92}, table~1, $\lambda\lambda6023$, 6275, 6340, 6460, 6537, 6578, are not  detected in \cite{winckel02}.
I have checked that $\lambda\lambda6023$ and 6340 do not appear in ESO spectra of the Red Rectangle.
The four other lines fall at  positions of telluric emission lines \cite{hanuschik03}.

Two other   bands in Scarrott  \emph{et al.} \cite{scarrott92} ($\lambda\lambda6395$, 6427) may correspond to lines in table~3 of \cite{winckel02}, but at significantly different peak wavelengths (6399 and 6421~\AA).
\section{Data analysis} \label{ana}
In this section and for each  Red Rectangle band of Table~\ref{tbl:1}, nebular and background spectra are compared in order to see whether or not the band has an atmospheric origin.
If it has a correspondence in the background, 'bg'  is reported in the last column of the table:
'bge' if it is an atmospheric emission line, 'bga' otherwise.
'bga' means that either the band is due to absorption by the atmosphere, or that it exists in the spectrum of HD44179 and belongs to the scattered starlight (in the atmosphere) component of the nebular spectrum (see the following paragraphs and Sect.~\ref{rrdibbg}).
An 'n'  in the last column of the table means that no straightforward indication of atmospheric origin is found, a '?' that the feature may not be real.

As in \cite{rrb} background spectra are re-scaled (magnified) so that the amplitude of their variations becomes comparable to those of the nebular spectra, in order to highlight atmospheric absorption features in the spectrum of the nebula.

A telluric emission line is an additive constant, added in the same way to all (background or nebular) spectra, independently of its continuum.
It is thus generally easily subtracted during the data reduction process.
Since the spectra will be considered before correction of atmospheric absorption, telluric emission lines must appear much larger in the re-scaled background spectra than in the spectra of the nebula.

Absorption by atmospheric molecules affects and modulates the continuum of each spectrum.
The resulting fluctuations are, in first approximation, in proportion of the continuum.
Therefore, especially if it is weak, the background should be re-scaled prior to a meaningful comparison with the spectrum of the nebula.
Atmospheric absorption features will then have similar amplitudes on both spectra, in contrast to telluric emission lines.
Re-scaling may also highlight small features in the spectrum of HD44179 which may be present in its light scattered in the atmosphere (see Sect.~\ref{dis}).

In \cite{rrb} I pointed out that the absorption spectrum of a complex molecule comprises 'valleys' and 'peaks', in a similar fashion as an emission spectrum:  absorption may thus be easily confused with emission.
It also appeared that correction for atmospheric absorption is far less obvious than subtraction of telluric emission lines is.
\subsection{$\lambda5768$} \label{l5768}
$\lambda5768$ is a weak, diffuse band first detected by Sarre  \emph{et al.} \cite{sarre95} on the blue edge of the $\lambda5800$ complex.
It begins at $\sim 5764\,\rm\AA$, peaks around 5768~\AA, and has a long red tail that ends between $5775$ and $5785\,\rm\AA$.

In ESO observations $11''$ from HD44179 it is seen from  $\sim s347$ to $\sim s410$, $17''$ on the sky, and from $\sim s324$ to $\sim s410$  (87 pixels, 23$''$) in the observation $6''$ from HD44179.
It is then too weak to be detected, although re-scaled backgrounds do present a similar feature  with minima at the same positions ($\sim 5760$ and $\sim5785\,\rm\AA$).

\begin{figure}[h]
\resizebox{\columnwidth }{!}{\includegraphics{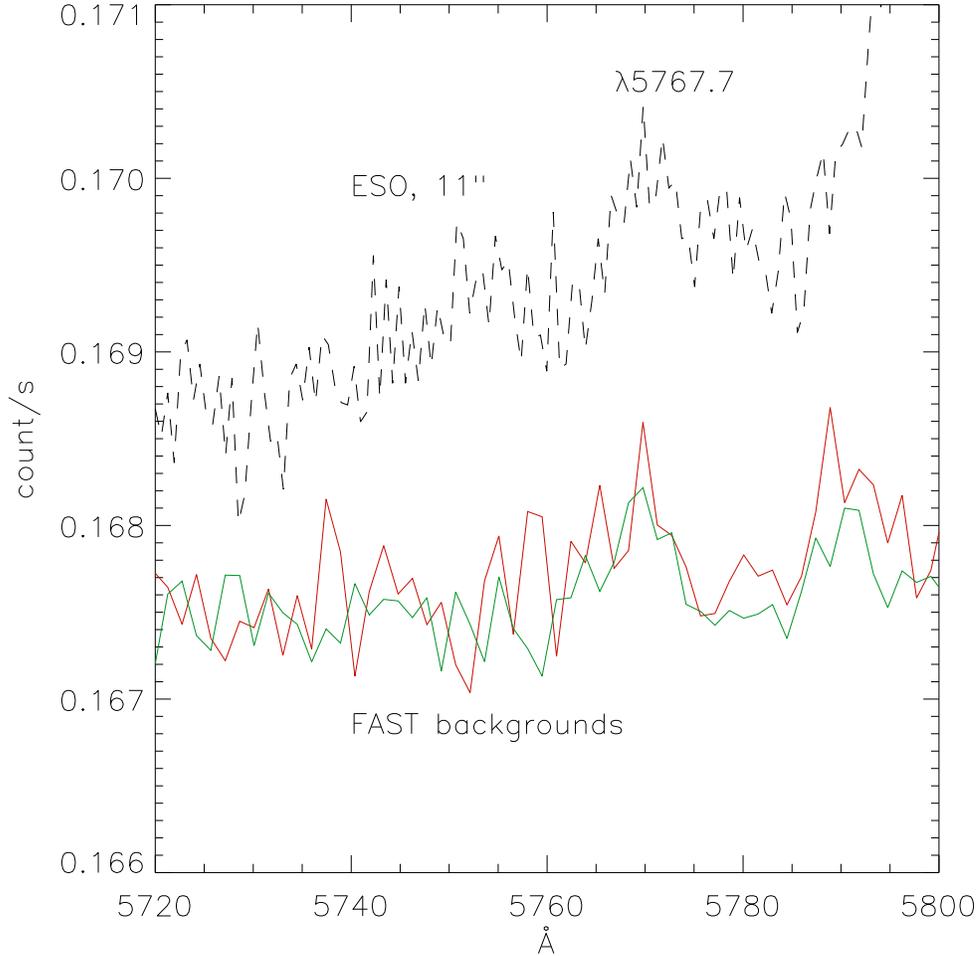}} 
\caption{ $\lambda5768$.
ESO observation $11''$ from HD44179, average of $s345$ to $s364$ in the nebula, and two FAST background spectra (averages of $s10$ to $s30$ and $s90$ to $s110$ on each side of the slit).
FAST backgrounds have been re-scaled for the comparison.
} 
\label{fig:fig1}
\end{figure}

This feature is well identified and more marked in the FAST backgrounds (Fig.~\ref{fig:fig1}), and therefore likely related to extinction by the atmosphere (and of 'bga' type).
\begin{figure}
\resizebox{\columnwidth }{!}{\includegraphics{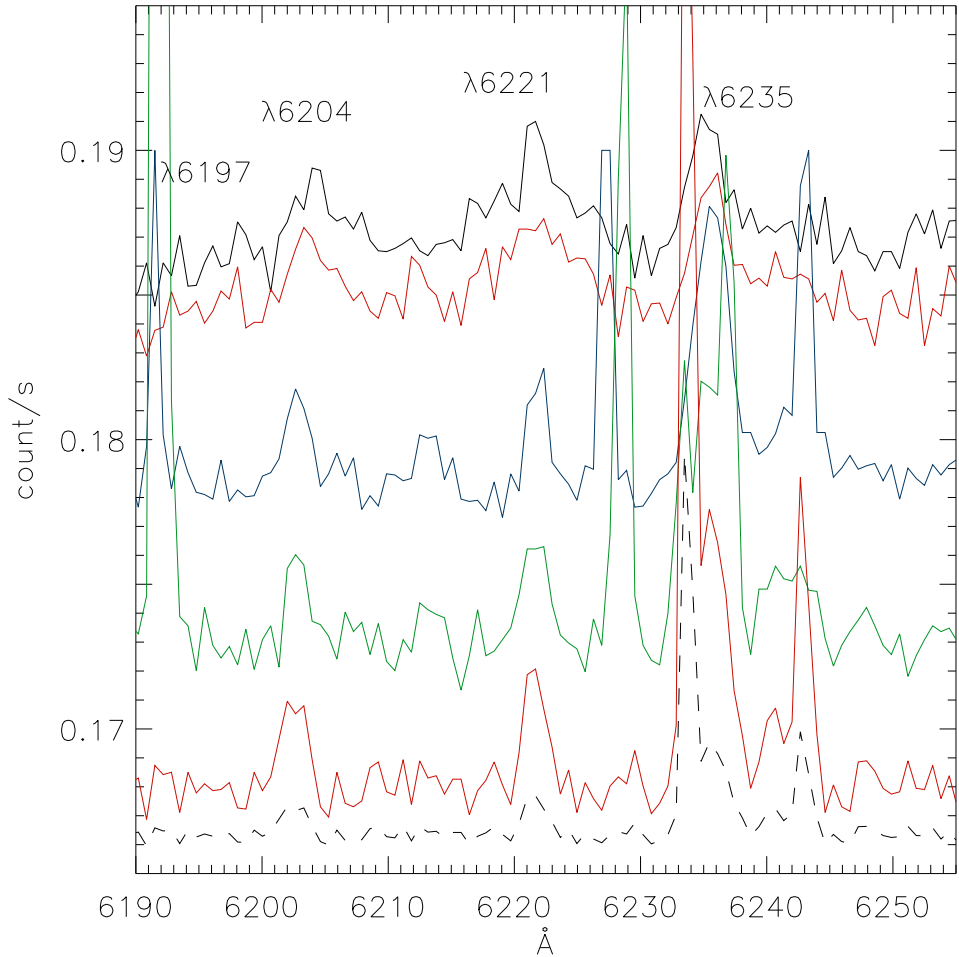}} 
\caption{Fine structure of the $\lambda6225$ diffuse band. 
The nebular spectrum (top spectrum), average of $s355$ to $s364$ along the slit, in the first ESO observations $11''$ north from HD44179, is compared to three re-scaled background spectra from same observation
(from top to bottom, averages of $s260$ to $s280$, in blue, $s460$ to $s480$, in green, $s520$ to $s540$, in red).
Background spectra are multiplied by three and arbitrarily shifted for sake of clarity.
Top red spectrum is the spectrum of the nebula (average of $s355$ to $s364$, no scaling) in the second ESO observation.
The bottom dashed spectrum is an example of  background ($s520$ to $s540$) shifted (+0.004~$count/s$) but  not re-scaled.
} 
\label{fig:fig2}
\end{figure}
\begin{figure}
\resizebox{\columnwidth }{!}{\includegraphics{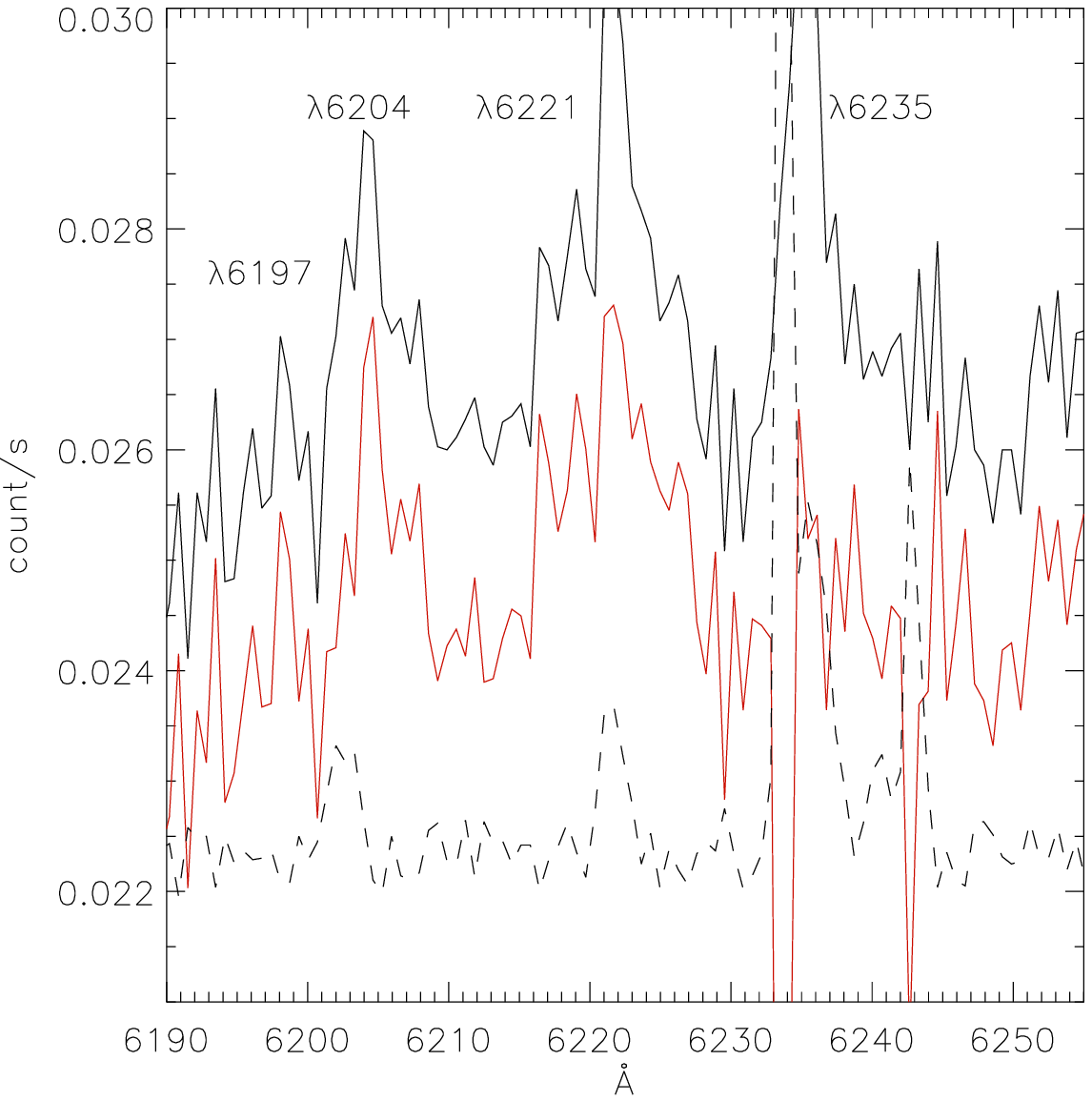}} 
\caption{Fine structure of the $\lambda6225$ diffuse band. 
Top spectrum is, as in Fig.~\ref{fig:fig2}, a nebular spectrum (average of $s355$ to $s364$), in the first ESO observations $11''$ north from HD44179, shifted by -0.1605~$count/s$.
Bottom dashed spectrum is a background spectrum (average of $s520$ to $s540$ offset by -0.14~$count/s$).
Middle red spectrum is the spectrum of the nebula minus the background.
Structures in the background clearly remain in the spectrum of the nebula after subtraction of the background, and will not be removed by normalization by a standard star.
In general  standard data-reduction routines are inefficient in correcting these effects of atmospheric extinction.
} 
\label{fig:fig2b}
\end{figure}
\subsection{$\lambda5799$ and $\lambda5853$} \label{l5800}
These two red-degraded Red Rectangle bands were the first to be related to DIBs ($\lambda5797$ and $\lambda5850$) \cite{fossey90}, and  are the strongest bands in the $\lambda5800$ complex.
The complex as a whole  is  proportionally similar at different distances from HD44179  (sect.~5.3 and figs.~11 and 12 in \cite{rrb}).
Two bands in the complex, $\lambda\lambda5912$ and 5937, are found in background spectra.

There is no indication that $\lambda5799$ and $\lambda5853$ are of atmospheric origin.
The only problem one may find with these bands is that they are detected over  $\sim27''$ in ESO  observations of the nebula $6''$ from HD44179, where the nebula is given to be $9''$ wide in \cite{winckel02}.
However, Van Winckel et al. give no indication on how  they derived the width.
\subsection{$\lambda6197$} \label{l6197}
This feature belongs to the $\lambda6225$ complex (next sub-section).
It  is not clear to me whether it is real or not.

On its red edge the band is delimited by the minimum  on the blue side of Red Rectangle band $\lambda6204$ (Fig.~\ref{fig:fig2}).

Towards the blue, however,  it is not possible to separate the feature from the continuum.
\subsection{The $\lambda6225$ complex} \label{l6225}
$\lambda6204$,  $\lambda6221$,  $\lambda 6235$ are sub-structures in Schmidt et al.'s $\lambda6225$ diffuse band.
The spectrum of the nebula $11''$ from HD44179 is, in this wavelength region, shown on  Fig.~\ref{fig:fig2} (top spectra).

Van Winckel \emph{et al.} \cite{winckel02} relates red-degraded $\lambda6204$,  $\lambda 6235$ to DIBs $\lambda 6203$ and $\lambda 6234$ of  the $6200\,\rm\AA$ DIB complex \cite{chlewicki87}.
DIB $\lambda6199$ is also close to the blue side of $\lambda6204$.

$\lambda6221$ has a symmetric profile, with DIBs $\lambda6216$ and $\lambda6226$ on each of its sides.

$\lambda6204$ and  $\lambda6221$ have clear counterparts in the background (Fig.~\ref{fig:fig2}) of ESO observations.
Since they have similar amplitudes on the nebular and re-scaled background spectra, they are not atmospheric emission features (they are of bga type).

$\lambda 6235$ also has a correspondence in the background (see also middle right plot of fig.~7 in \cite{rrb}).
The line is much larger on the re-scaled background spectra  than on the nebular ones, which indicates it is an atmospheric emission line.
It does  fall at the exact position of an emission line identified in the La Silla-Paranal night sky spectrum (fig.~23 in \cite{hanuschik03}).

Fig.~\ref{fig:fig2b} shows that subtraction of the background largely removes $\lambda6235$ but has no significant effect on $\lambda\lambda 6204,\,6221$.
These bga features still remain in the Van Winckel et al. \cite{winckel02} spectra after complete data-reduction; they are thus not corrected by standard routines.

These observations confirm the finding in \cite{rrb}, that the structure of the nebula's spectrum in the $6200\,\rm\AA$ region is determined by atmospheric extinction.
\subsection{ $\lambda6446$} \label{l6446}
\begin{figure}
\resizebox{\columnwidth }{!}{\includegraphics{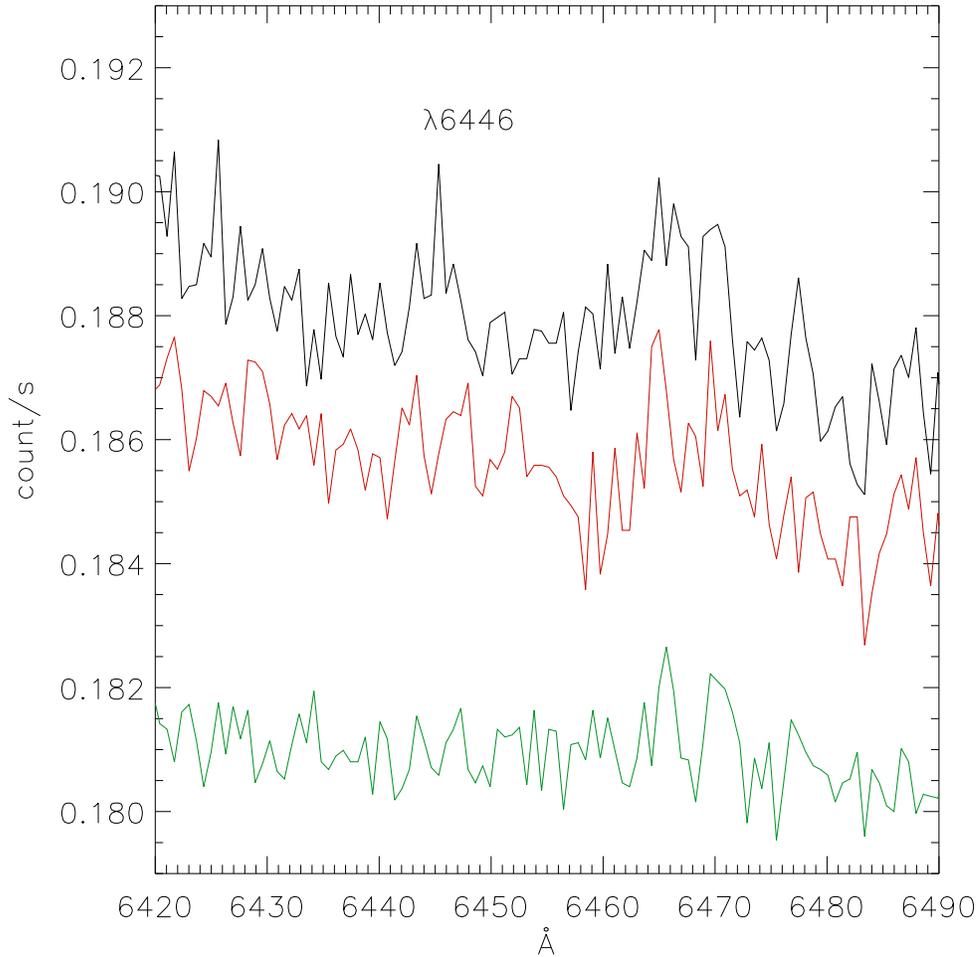}} 
\caption{$\lambda6446$.
Top spectrum is the average of $s355$ to $s364$ of the first ESO observation of the nebula, $11''$ from HD44179.
In green, the average of  $s340$ to $s348$ (shifted by +0.01),  same observation.
Red spectrum is the average of $s353$ to $s361$ in the second, similar, ESO observation.
All three spectra are in the nebula.
The group of lines close to 6465~\AA\ are night sky emission lines \cite{hanuschik03} also present in background spectra (not represented here). 
} 
\label{fig:fig3}
\end{figure}
$\lambda6446$  is a small feature which appears on the average of a few  spectra ($s355$ to $s364$, Fig.~\ref{fig:fig3}) within the nebula, in the first ESO observation $11''$ from HD44179.
However it  cannot be distinguished from the continuum in the individual spectra of this observation, or in other averaged spectra (e.g., from s340 to s348, Fig. 3).
It is also absent from the average spectrum of the second identical ESO observation.
It may therefore be an artifact.
\begin{figure}
\resizebox{\columnwidth }{!}{\includegraphics{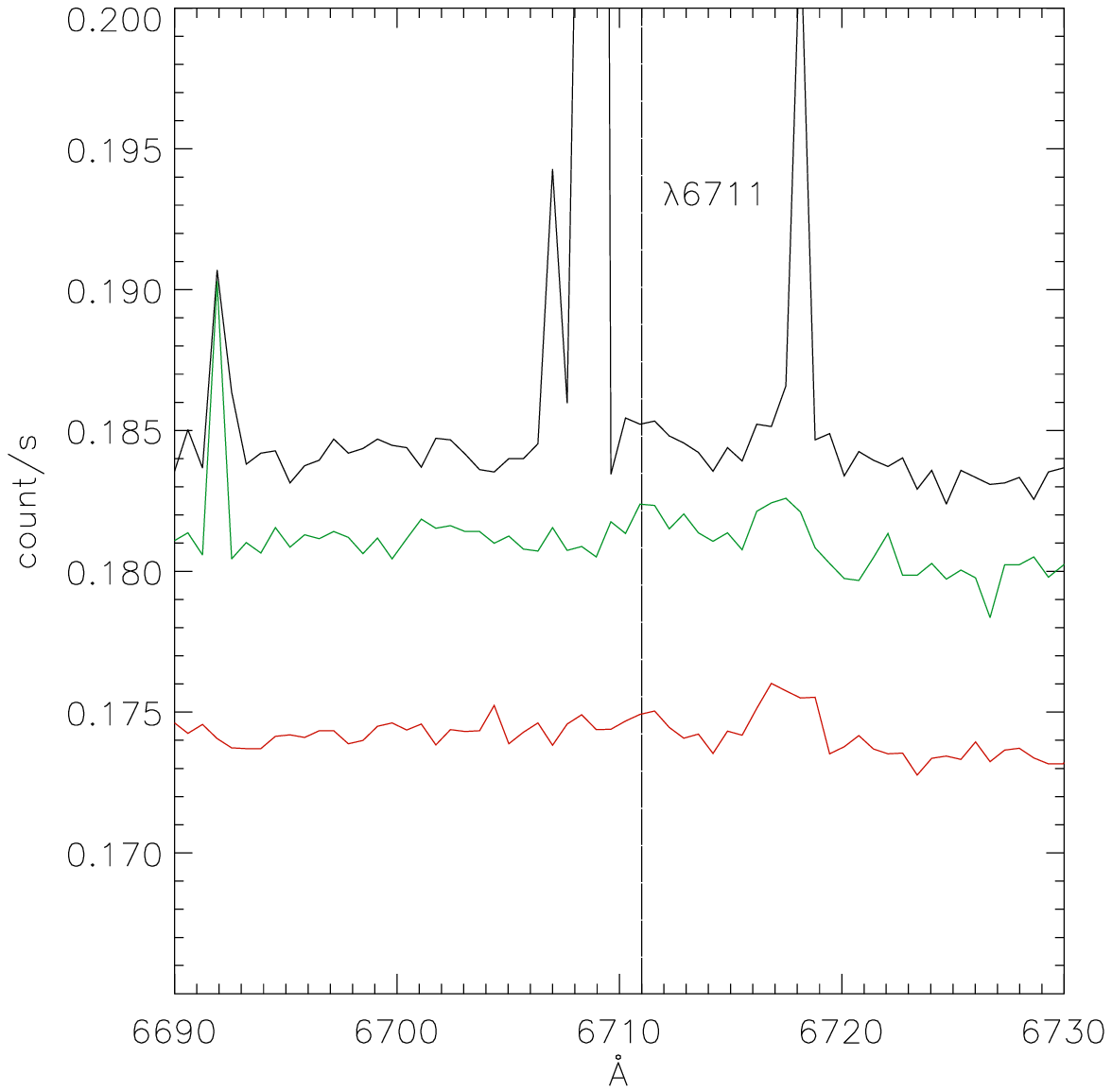}} 
\caption{ $\lambda6711$.
In black, average spectrum of $s349$ to $s358$, in the nebula,  first ESO observation, $11''$ from HD44179.
Green is the average of  $s359$ to $s370$, still in the nebula, same observation.
Red spectrum is the average of $s340$ to $s360$ (in the nebula) in the second, identical, ESO exposure. 
The bump centered at 6717~\AA\ in the two later spectra is an emission sky line listed in \cite{hanuschik03}.
} 
\label{fig:fig4}
\end{figure}
\subsection{ $\lambda6615$} \label{l6615}
$\lambda6615$ is a more precise determination of Schmidt et al.'s $\lambda6617$ band.
In ESO observations $11''$ from HD44179 it is $\sim 10\,\rm\AA$ wide (from $\sim6610$ to $\sim 6620\,\rm\AA$).

It is observed on a slightly larger extent than found for the nebula (\cite{rrb}, sect.~5.2), and the background seems to reproduce its fine structure.
These observations are however at the limit of the error margin, thus can not be used to discuss rigorously the  origin of the band.
\subsection{ $\lambda6711$} \label{l6711}
As for $\lambda6197$ and $\lambda6446$  the existence of this last feature, from an analysis of individual spectra in ESO observations $11''$ from HD44179, does not seem to me obvious.

There is  a spike (Fig.~\ref{fig:fig4}) in a few spectra (from $s349$ to $s358$) of the first ESO observation $11''$ from HD44179.
This spike, probably a cosmic ray, is absent from all other spectra as well as in the second identical observation of the nebula (Fig.~\ref{fig:fig4}).
There may be a slight enhancement of the continuum  around 6711~\AA\ on Fig.~\ref{fig:fig4} but not significant enough to be considered as evidence, as it is done in \cite{winckel02}, for an emision line.
\section{Discussion} \label{dis}
\subsection{The Red Rectangle bands} \label{rrnat}
Reserves should be made on the reality of three ($\lambda\lambda6197$, 6445, 6711) out of the nine Van Winckel et al. bands associated to DIBs.
$\lambda\lambda6445$ and 6711 seem to be accidental artifacts in a few spectra of one of the  observations $11''$ from HD44179.
$\lambda6197$ may be caused by the minimum on $\lambda6204$'s blue side.

Four other bands ($\lambda\lambda5768$, 6204, 6221, 6235) have a correspondence in the background, which I believe is not accidental.
One ($\lambda6235$) is a telluric emission line listed in \cite{hanuschik03}.
The three others (two of which belong to the $\lambda6225$ complex), unless they are related to light from HD44179 scattered in the atmosphere  (Sect.~\ref{rrdibbg}), must be due to absorption in the atmosphere.

Sect.~\ref{rrselect} has shown that similar bias exist in other data-sets:
over 50~\% of the bands  found in Scarrott \emph{et al.} \cite{scarrott92} outside the 5800~\AA\ region are not detected in Van Winckel \emph{et al.} \cite{winckel02}, and are probably either artifacts or due to the atmosphere.

The three remaining bands in Table~\ref{tbl:1}, among which two are in the $\lambda5800$ complex,  may arise in the nebula, although this remains to be proved.

This study thus confirms the importance of the atmosphere, especially in the 6200~\AA\ region, in the observed spectrum of the Red Rectangle.
\subsection{The relationship between DIBs and Red Rectangle bands} \label{rrdib}
DIBs may now be found in any part of the visible spectrum, and it will in general be easy to find  DIBs in the vicinity of each Red Rectangle band (see fourth column of Table~\ref{tbl:1}).

On the one hand, the uncertainty on the Red Rectangle bands, their relatively low (compared to S/N ratios achieved in DIB observations) level of detection, show how careful one must be when associating DIBs and Red Rectangle bands.

On the other hand DIB/Red Rectangle band associations do not rely only on wavelength coincidences.
Other criteria, as the importance of the bands in each spectrum, have to be taken into account.
In the  5800~\AA\ region for instance, it would indeed be remarkable that two sets of prominent features, in the spectra of reddened stars and in the spectrum of a nebula, can coincide and be independent.

There might thus be less DIB/Red Rectangle band true associations than previously though.
In addition, the difficulties  raised by the relative positions of these DIBs and Red Rectangle bands on the spectrum, and the objections in \cite{glinski02}, could be overcome if the Red Rectangle spectrum was an absorption spectrum:
it is then the valleys on the steep sides of the Red Rectangle bands in the nebular spectra which have to coincide with DIBs, as observed.
\subsection{Red Rectangle bands in background spectra} \label{rrdibbg}
Red Rectangle bands $\lambda5768$ and $\lambda\lambda6204$, 6221, 6235 in the $\lambda6225$ complex,  are found in background spectra.

The presence of DIBs $\lambda\lambda6234$ and 6237 on the sides of $\lambda6235$ may be a coincidence since the band is at the exact position of a telluric emission line.

The presence in the background of the other bands and a relation to DIBs is more difficult to understand, but may be justified by either of the following possibilities.

These Red Rectangle bands can be due to an incomplete removal of atmospheric absorption \cite{rrb}.
If they are related to DIBs,  it will necessarily be concluded that the corresponding DIBs are also atmospheric absorption bands, and have not been removed during the data reduction process.
Next section shows that this hypothesis, as unexpected as it can be, can not be ruled out.

This hypothesis could explain the  complexity of the DIB spectrum, but would also mean that some DIBs need to have, in addition to an interstellar cloud, the atmosphere on the line of sight.
Should it be the case, the relative roles of the interstellar medium and of the atmosphere, in the formation of these DIBs, need to be elucidated.
It cannot concern all DIBs since some are detected with high redshifts in other galaxies  (for instance \cite{ehren02} for the Magellanic Clouds, \cite{sollerman05} for NGC~1448).

Another possibility is that DIBs, although they are not observed toward  HD44179,  exist at a low level  in the spectrum of the star, and are detected in the light from HD44179 scattered by the nebula and/or the atmosphere. 
This hypothesis would justify that the observed spectrum of the nebula is an absorption spectrum, and its relationship to the DIBs.

Alternatively, Red Rectangle bands due to the atmosphere may not be related to DIBs.

Space observations of the Red Rectangle nebula and of reddened stars with the Hubble Space Telescope (HST)  will provide the most straightforward answer to these hypotheses.
\subsection{The $\lambda6200$ DIB complex} \label{dib6200}
As the 5800~\AA\ region, the 6200~\AA\ region (from $\sim6150$ to $\sim6250\,\rm\AA$) concentrates many features whose existence is doubtless, and coincide in the DIB and Red Rectangle spectra.

In \cite{rrb} the diffuse $\lambda6225$ Schmidt et al. Red Rectangle band was attributed to atmospheric absorption, and more specifically to absorption by O$_4$ on its red side (sect.~6.3 and fig.~14 in \cite{rrb}).
Sect.~\ref{l6225} indicates that its fine structure may also be due to the atmosphere.
A  link between the Red Rectangle band and the $\lambda6200$ DIB complex would thus imply that the latter results from atmospheric absorption.

\begin{figure}[h]
\resizebox{\columnwidth }{!}{\includegraphics{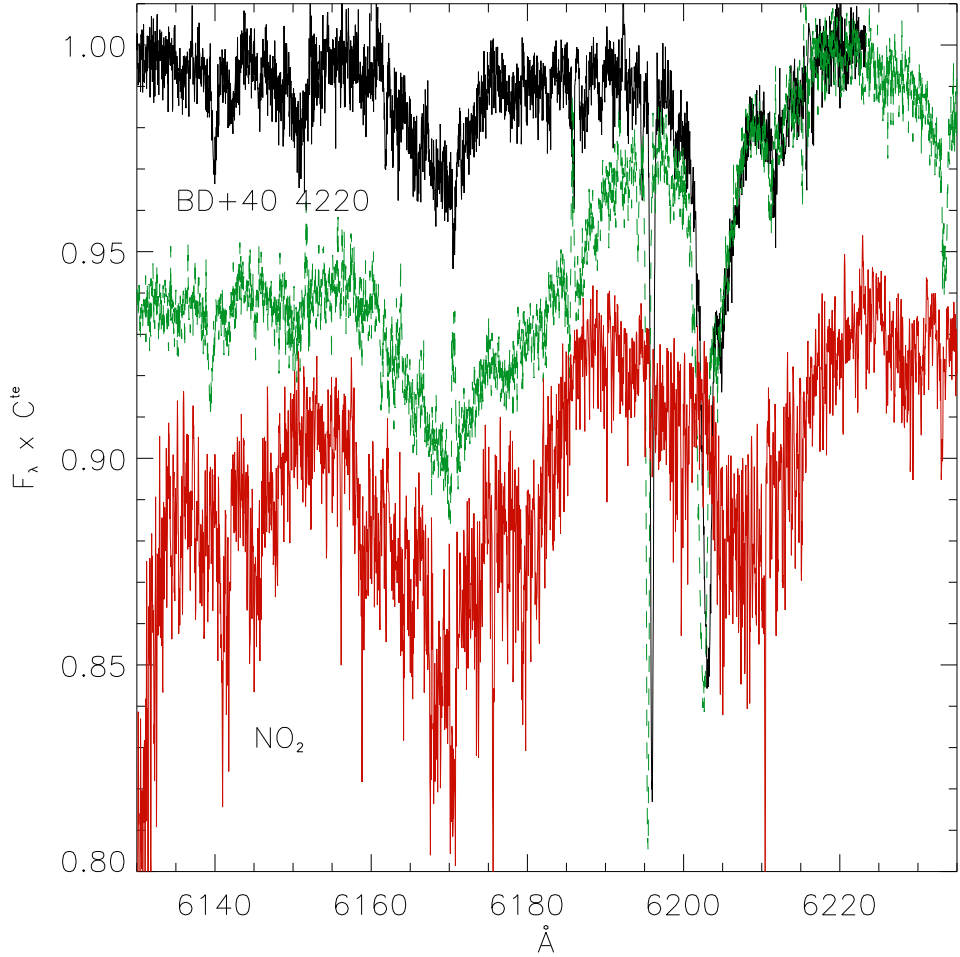}} 
\caption{
The raw (before correction for atmospheric extinction) spectrum of BD+40$^{\circ}$4220 (resolution=0.03~\AA) in green (courtesy of G.~Galazutdinov) is compared to the absorption spectrum of NO$_2$ (in red, resolution = 0.01~\AA) in the 6200~\AA\ region.
The top black spectrum is  BD+40$^{\circ}$4220's after data reduction (fig.~14 in \cite{gala00}).
} 
\label{fig:fig5}
\end{figure}

Fig.~\ref{fig:fig5} compares the raw spectrum (before corrections for atmospheric extinction, radial velocity and normalization to the continuum) of  BD+40$^{\circ}$4220 (green spectrum of the figure) observed by G.~Galazutdinov, and A.~Jenouvrier's laboratory spectrum of NO$_2$ (in red), one among the most important and complex molecules in the atmosphere.

There are evident similarities between the spectra.
Low frequency variations are common to both.
The broad DIB $\lambda6177$, reported in earlier papers \cite{chlewicki87,lebertre93} and seen here with much better resolution, is a clear, recognizable feature of the spectrum of NO$_2$.
Narrower DIBs in the  \cite{gala00} catalog,  $\lambda6139.94$, $\lambda6145.65$ $\lambda6158.54$, $\lambda6167.84$, $\lambda6220.82$, $\lambda6223.56$, $\lambda6226.3$ also have  counterparts in the spectrum of NO$_2$.
Differences of position of the features' minima between the spectra are insignificant, considering the complexity of NO$_2$ absorption (which is not resolved in either observations), and the different observational conditions of the two spectra. 

The spectra also have differences, mainly the two sharp, strong absorption lines  (DIBs) at 6196 and 6203~\AA\ (on the blue side of  Red Rectangle band $\lambda6204$),  noticeably absent from the NO$_2$ spectrum, and two small absorption features at 6151~\AA\ and 6186~\AA.

I found no straightforward identification for  these lines.
DIBs $\lambda\lambda6196$ and 6204 were detected in NGC1448 \cite{sollerman05} with a redshift of $\sim24\,\rm\AA$ ($\sim1160$~km/s), which implies an interstellar origin.

On the right hand side of the plot,  the structure  between 6210 and 6230~\AA\ in the spectrum of NO$_2$ (in the spectrum of BD+40$^{\circ}$4220 as well) coincides with  Red Rectangle band $\lambda6221$ (Fig.~\ref{fig:fig2}).
Attribution of the diffuse $\lambda 6225$ Red Rectangle band to O$_4$'s absorption on its red side, and of part of its sub-structure to NO$_2$, would mean that several atmospheric molecules may contribute to the Red Rectangle bands in a given wavelength region.

The DIB spectrum (in black, see also figs.~14 and 15 in \cite{gala00}), obtained after data reduction, does not differ much from the raw spectrum except that the continuum is normalized  and redshifted (by $1\,\rm\AA\sim 30$~km/s) to correct for the radial velocity of the interstellar cloud on the line of sight and for the motion of the earth, assuming that absorption  is all interstellar.
Normalization of the continuum has rubbed part of the similarity with the NO$_2$ spectrum, especially the steep decrease between 6170~\AA\ and 6190~\AA.
The broad DIB absorption centered close to 6170~\AA, as well as smaller features are still present in the DIB spectrum.

It is  evident that, if NO$_2$ determines the structure of the raw spectrum in this wavelength region, the data-reduction process did not remove all effects of atmospheric absorption.
Some absorption bands which have been attributed to absorption in the interstellar cloud, are due to atmospheric NO$_2$ (possibly also to other molecules in the atmosphere).
\section{Conclusion} \label{conc}
This study has questioned the origin of  Red Rectangle bands which have been considered as DIBs in emission.
The presence  of some bands in background spectra questions the nature of the Red Rectangle spectrum (emission or absorption, interstellar or due to the atmosphere?), as it is observed from earth, the  Red Rectangle band/DIB relationships, and the nature of some DIBs.

I have also presented observations which show the remarkable similarity between the DIB spectrum of a reddened star and a laboratory spectrum of NO$_2$, in the $[6100,\,6240]$~\AA\ wavelength range.
If part of DIB complex $\lambda6200$  is indeed due to atmospheric absorption,  its observation in the spectrum of reddened stars needs, in addition to an interstellar cloud, to have the atmosphere on the line of sight.
In this case, the respective roles of the interstellar cloud and the atmosphere, in the observation of this DIB, need to be elucidated.

Observed DIBs and Red Rectangle spectra may therefore be caused by different kinds of extinction phenomena, interstellar but also atmospheric, which render the interpretation of these ground-based observations particularly difficult and complex.
To separate atmospheric and interstellar effects in the DIBs and Red Rectangle spectra, the easiest and most reliable way would be to obtain spectra  from outside the atmosphere, with the HST for instance (to this date, no such observational program has been performed). 
\section*{Acknowledgments}
I am highly indebted to  G.~Galazutdinov and to A.~Jenouvrier for allowing me to use  their observations of  BD+40$^{\circ}$4220 and NO$_2$ in this publication.
I also wish to thank anonymous referees for their careful reading of the manuscript and useful comments.

This research has used observations made with ESO NTT Telescope at  La Silla-Paranal  Observatory under programme ID 60.c-0473.

{}

\end{document}